# MathPartner Computer Algebra
G. I. Malaschonok

**Abstract** In this paper, we describe the general characteristics of the MathPartner computer algebra system (CAS) and its Mathpar programming language. The MathPartner can be used for scientific and engineering calculations, as well as in secondary schools and higher education institutions. It allows one to carry out both simple calculations (acting as a scientific calculator) and complex calculations with large-scale mathematical objects. The Mathpar is a procedural language that supports a large number of elementary and special functions, as well as matrix and polynomial operators. This service allows one to build function images and animate them. The MathPartner also makes it possible to solve some symbolic computation problems on supercomputers with distributed memory. We highlight the main differences of the MathPartner from other CASs and describe the Mathpar language along with the user service provided.

## 1. INTRODUCTION

There are several generations in the history of computer algebra systems (CASs). The first generation dates back to the 1960s–1970s.

Graphics-oriented systems, which appeared in the early 1990s, constitute the second CAS generation; as an example we can point to the Mathematica and Maple commercial systems, which are still the most popular CAS instruments.

The third generation is associated with cloud-oriented CASs like, for example, MathPartner, Sage, Wolfram Alpha, and Wolfram Cloud, which appeared in the 2010s. These systems are now becoming increasingly popular.

A CAS is a software product characterized by: algebraic algorithms employed, software technologies supported, and application domain intended.

The history of constructive algebra and algebraic algorithms extends back over more than a century. In turn, current investigations in this field depend heavily on the capabilities of presently-available CASs, as well as on the characteristics of the applied problems that can be solved within a reasonable time.

It is interesting to trace how the CAS application areas were changed. Initially, the main CAS application area covered symbolic computations in theoretical physics and celestial mechanics. One of the first CASs was the Schoonschip, developed by Martinus J.G. Veltman, a Dutch physicist, in the 1960s during his work at the Stanford laboratory. This system was put into use by Gerardus't Hooft, a student of Veltman. Hooft proved that non-Abelian gauge theories can be renormalized. In 1999, Veltman and Hooft shared the Nobel Prize in Physics "for elucidating the quantum structure of electroweak interactions" [1]. A descendant of the Schoonschip was the FORM system developed by Jos Vermaseren from the Dutch National Institute for Subatomic Physics (Nikhef) [2]. Its development began in 1984. To learn more about the CAS history, one may refer to [3, 4].

The advent of the second CAS generation expanded the CAS application area considerably. These systems became popular among many teachers and students of engineering and natural science disciplines, as well as among many engineers and scientists. Wolfram Research Inc. began to adapt the Mathematica system to secondary schools. The documentation on the Maple and Mathematica systems can be found at www.maplesoft.com and www.wolfram.com, respectively.

And yet, among those who study and employ mathematics, the percentage of CAS users still remains low, which is mainly because these systems are difficult to master and require purchasing a license.

With cloud CASs, these obstacles can be overcome. Let us point to their basic features.

– Cloud systems can be accessed from any computer connected to the Internet.

– They provide users with advanced services, including a workbook with a user-friendly interface.

– A fundamental difference with the new-generation CASs is the possibility of solving the problems that cannot be solved on a personal computer. This can be done through the access to parallel software packages installed on a computer cluster with distributed memory. The user does not need to worry about computing resources, since the cloud system is able to undertake the control of the computational process.

– An important feature is the support of the RESTFul protocol for data communication with other Internet-connected devices. Computational results can be obtained by the cloud CAS independently (without human assistance).

– Mathematical calculations in the form of the text files containing program code can be stored in shared libraries.

– Any mathematical problem can be solved by uploading such a text file from a shared library to the cloud workbook. Mathematics textbooks and reference books can be complemented with the libraries containing text files with the corresponding algorithms designed for computations in the cloud workbook.

– A cloud CAS can provide an educational service, including a library of tests, automatic test checking, and storing test results in the student's record book.

The MathPartner system was one of the first cloud CASs [5].

This project has been started in 2002 by a Tambov working group on computer algebra. The basis of the MathPartner software package are the algorithms developed at the Lvov, Kiev, and Tambov State Universities (more details are available on the Internet [6]).

Over the years, the development of the algorithms and programs for the MathPartner project involved: A.A. Betin, I.A. Borisov, Yu.D. Valeev, E.V. Dubovitskii, A.M. Dobychin, M.S. Zuev, D.S. Ivashov, E.A. Il'chenko, S.A. Kireev, V.A. Korabel'nikov, V.N. Kazakov, A.O. Lapaev, G.I. Malaschonok (head), O.N. Pereslavtseva, A.G. Pozdnikin, M.A. Rybakov, O.A. Sazhneva, R.A. Smirnov, M.V. Starov, S.M. Tararova, S.A. Khvorov, D.I. Shlyapin, A.S. Shcherbinin, and Yu.Yu. Yurin. Since 2011, the project is being developed by the Mathparca.

Let us highlight the main differences of the MathPartner from other CASs.

## 2. GENERAL INFORMATION ABOUT THE MATHPARTNER

The MathPartner is a universal cloud-oriented system for symbolic computations, which is based on some open-source software packages. The MathPartner kernel is a mathematical software package written in Java.

The user language is called Mathpar. Having the syntax that maximally resembles the standard mathematical notation, this language can be regarded as a dialect of the LaTeX language, which is extended with operators for symbolic computations, image visualization, and environment setting.

Having entered the MathPartner website, each user is provided with a personal workbook. This workbook has an active window through which the user can enter certain text (which is enclosed in double quotes) and mathematical expressions written in the Mathpar language.

The user can create new active windows in his or her workbook and run programs in any of the windows independently. In this case, all the windows use the same space of variables.

The user can save (to his or her computer) and, if needed, upload (to the workbook) original text files in the Mathpar language. The MathPartner allows the user to get LaTeX and PDF files and to save them.

In addition, the MathPartner makes it possible to create and store 2D and 3D images and to animate them. The user can build images of tabulated functions, select certain regions on these images, mark points, construct approximations, make notations, and so on.

The Mathpar is a procedural language and can be used to write user programs. The MathPartner system has quite a wide area of application. It can be used as a simple calculator, scientific calculator, software system for symbolic computations, and interface to a computer cluster with distributed memory.

Moreover, the MathPartner allows one to import and export large-scale mathematical objects such as functional or numeric matrices and complex composite functions or polynomials with a large number of monomials. There is no need to fully upload such objects to the workbook: they can be stored (as files) on the user computer. To import and export large-scale mathematical objects, the following Mathpar operators are used: fromFile() and toFile().

The Mathpar language also contains some operators that enable symbolic computations on distributed-memory supercomputers. These operators can be used by the registered users.

## 3. ABOUT THE MATHPAR LANGUAGE

The MathPartner allows the users to upload and run programs written in the Mathpar language. Below is the general description of the Mathpar.

– The Mathpar language employs common LaTeX symbols and contains assignment operators, calculation operators, control operators, environment operators, and plot operators. The system uses simple text files with the text written in the Mathpar language, which can be created, edited, and saved just like standard text files encoded in UTF-8.

– The Mathpar defines environment operators for setting the space of mathematical objects. The environment determines the underlying set of numbers, algebraic operations in this set, names of variables, and some constants.

– A Mathpar program and computational results can be represented in the form of an editable source text or a PDF image. The user can change the view in the workbook.

– The Mathpar defines a group of operators for solving computational problems on a computer cluster with distributed memory.

The Mathpar is a procedural language. To learn more about the Mathpar, one can visit the help page on the project website, download the Mathpar language guide from the website, or see [7–11]. Below, we describe only its basic features.

## 4. SETTING THE COMPUTATIONAL ENVIRONMENT

The environment defines the space wherein computations are carried out. It allows one to specify the underlying set of numbers, algebraic operations in this set, names of variables, and some constants.

By default, the space $R64[x, y, z, t]$ is set. It is a space of four variables (ranked as $x < y < z < t$) over the set of approximate real numbers, which are stored in a 64-bit machine word (type "double" for C and Java).

To change the environment, it is required to execute the command for setting a new environment, for instance, SPACE = $Q[x]$ or SPACE = $Z[c, b, a]$.

The user can select one of the following number sets: Z; Zp=$Z/pZ$ (p > 2, where p is a MOD environment constant); Zp32=$Z/pZ$ (p < $2^{31}$, where p is a MOD32 environment constant); Z64=$\{z \in Z : -2^{63} \leq z < 2^{63}\}$; Q; R is a set of approximate real numbers for which the number of digits in the mantissa is determined by the ACCURACY environment constant, while the machine epsilon is determined by the MachineEpsilonR constant; R64 is a set of approximate real numbers (52 bits represent the mantissa and 11 bits represent the exponent) for which the machine epsilon is determined by the MachineEpsilonR64 constant; and R128 is a set of approximate real numbers (128 bit). Another eight number sets---CZ, CZp, CZp32, CZ64, CQ,

C, C64, and C128---are obtained by complexifying the above eight sets. More information about these and other environment constants can be found in the user manual.

Such an explicit setting by the user of a numerical environment for symbolic calculations is due to the increased requirements for modern CAS. It should be noted that most CASs fix only one environment that cannot be changed by the user.

The majority of users will be satisfied with the environment set by default; it, however, will not be sufficient for many other users.

For instance, to carry out approximate computations with a given number of exact digits in arithmetic operations and a given value of the "machine epsilon", one can set the space R and select the corresponding constants.

If it is required to carry out computations over a set of complex numbers, then the space C should be selected. If only rational computations need to be carried out, then the space Q will be an appropriate choice.

When there is a need to prepare some tasks for pupils who are familiar only with integers or only with fractional numbers, the spaces Z or Q should be set, respectively.
Without setting a proper environment, the user cannot carry out computations, whether it be in finite fields, tropical mathematics, etc.

When an integer has more than 300 decimal digits, it cannot be written in an approximate form (in a variable of the double data type) because of the overflow. However, it can be written as a number of the R128 type, since this data type provides 64 extra bits for the exponent. Moreover, if necessary, the user can execute standard mathematical operations over such numbers.

All this extends the capabilities of the CAS, which becomes a flexible and convenient tool.

An important feature of the MathPartner system is dividing all symbolic variables into two sets: a set of symbols and a set of variables defined in the SPACE and selected by the user.

If the user has specified variables in statement SPACE, then the compiler will generate polynomials and polynomial fractions from these variables, just like some standard objects. For these objects, efficient computational algorithms will be applied.

# 5. MAIN CLASSES OF OPERATORS

## 5.1. Mathematical Symbols and Functions

The symbols $\infty$, $e$, $i$, $\pi$, and $\oslash$ are defined, which are written just like in the LaTeX language ($e$ and $i$ are written with an additional symbol "\ ").

The functions defined in the Mathpar language are described in the user manual (see also the help page on the official website). For the convenience of the user, functions can be selected and transferred to the book from a pull-down side menu.

## 5.2. Standard Sets of Operators

Like other CAS languages, the Mathpar uses the following standard sets of operators:
 – Operators for numbers.
 – Operators for integers (particularly).
 – Operators for fractions and rational functions.
 – Operators for function evaluation in probability theory and mathematical statistics, which are defined for continuous random variables, discrete random variables, or samples.
 – Boolean operators for the set {true, false}={1, 0} and matrices over it.
 – Comparison operators for numbers.
 – Set theory operators for the algebra of subsets of real numbers: $\cap, \cup, , , \triangle$, and a complement denoted by ´.

– Operators for constructing random mathematical objects: numbers, polynomials, and matrices over numbers or polynomials.

## 5.3. Operators over Polynomials in Many Variables

This is a principal set of operators for any CAS. In the MathPartner, the representation of polynomials in many variables is selected so as to ensure the efficiency of computations with large sparse polynomials.

It should be noted that, for polynomials, an F4 algorithm is implemented to compute the Groebner basis $\mathbf{groebner}(f_1, f_2, .., f_c)$ of the ideal generated by a finite set of polynomials ($f_1, f_2, .., f_c$).

There is also an algorithm $\mathbf{reduceByGB}(g, [f_1, f_2, .., f_c])$ for reducing the polynomial $g$ by using the polynomials $f_1, f_2, .., f_c$. If these polynomials constitute the Groebner basis, then the algorithm evaluates the reduction modulo the ideal generated by them.

If the variety of solutions for a system of nonlinear algebraic equations $f_1 = 0, f_2 = 0, .., f_c = 0$ has a zero dimension, then this system can be solved by the operator $\mathbf{solveNAE}(f_1, f_2, .., f_c)$.

The operator $\mathbf{solve}(p(x))$ calculates the roots of a polynomial $p(x)$. The set wherein the roots are found depends on the space (SPACE) selected. If the space is defined over real numbers, then all real roots are found; if it is defined over complex numbers, then all complex roots are found; in the other cases, the roots containing radical signs and literal parameters are found (if possible).

Complexity estimates of the polynomial algorithms can be found in [12].
The number of publications [6] concerning the algorithms employed in the MathPartner project exceeds two hundred papers, which is why we refer here only to the selected ones.

## 5.4. Matrices and Vectors

The set of symbolic matrix operators holds a central position in the MathPartner system. Experiments showed that, in terms of computational efficiency and functionality, these operators, in certain cases, are superior to those used in other CASs.

Take, for example, a 12×12 matrix in which the elements denoted by $x$ are independent symbols and the other elements are zero:

$$\begin{bmatrix} x & x & x & 0 & 0 & 0 & x & x & x & 0 & 0 & 0 \\ 0 & x & x & x & x & 0 & 0 & x & x & x & x & 0 \\ 0 & 0 & 0 & x & x & x & 0 & 0 & 0 & x & x & x \\ x & x & x & 0 & 0 & 0 & 0 & 0 & 0 & 0 & 0 & 0 \\ 0 & x & x & x & x & 0 & 0 & 0 & 0 & 0 & 0 & 0 \\ 0 & 0 & 0 & x & x & x & 0 & 0 & 0 & x & x & x \\ x & x & x & 0 & 0 & 0 & x & x & x & 0 & 0 & 0 \\ x & x & x & x & 0 & 0 & x & x & x & x & 0 & 0 \\ 0 & 0 & 0 & x & x & x & 0 & 0 & 0 & x & x & x \\ 0 & 0 & 0 & 0 & 0 & 0 & x & x & x & 0 & 0 & 0 \\ 0 & 0 & 0 & 0 & 0 & 0 & 0 & x & x & x & x & 0 \\ 0 & 0 & 0 & x & x & x & 0 & 0 & 0 & x & x & x \end{bmatrix}.$$

The determinant of this matrix can be evaluated in the MathPartner, but not in the Mathematica and not in the Maple.

Below are the matrix operators of the Mathpar language.
– $\mathbf{kernel}()$ is the kernel of the operator (null space).
– $\mathbf{transpose}(A)$ or $A\{T\}$ is the transposition.

- **conjugate**(*A*) or $A^{\ast}$ is the conjugation.
- **toEchelonForm**() is the echelon form.
- det() is the determinant.
- rank() is the rank.
- **inverse**(*A*) or A^{-1} is the inverse of a matrix.
- **adjoint**(*A*) or A^{\star} is the adjoint of a matrix.
- **genInverse**(*A*) or A^{+} is the Moore–Penrose pseudoinverse.
- **charPolynom**() is the characteristic polynomial.
- **closure**(*A*) or A^{\times} is the closure, i.e., the sum $I + A + A^2 + A^3 + \ldots$.
- **SimplexMax**() and **SimplexMin**() are used to solve linear programming problems. For example, solves the system of inequalities $Ax \leq b$ so that the scalar product $c^T x$ is minimal. Note that here one can obtain both rational solutions and solutions with an accuracy required.
- **LSUWMdet(A)** is the triangular LSU factorization of a matrix, which yields five matrices L,S,U,W,M. The product LSU corresponds to the original matrix A. Here, $L$ is a lower triangular matrix, $U$ is an upper triangular matrix, S is a permutation matrix multiplied by a diagonal matrix. More over P=det^(-2)*WSM is the inverse matrix for the matrix A. If the matrix A is not invertible then P is a pseudo inverse: P=PAP and A=APA.
- **BruhatDecomposition**() is the Bruhat decomposition of a matrix, which yields three matrices $[V, w, U]$, where $V$ and $U$ are upper triangular matrices and $w$ is a permutation matrix multiplied by a diagonal matrix (the product $VwU$ corresponds to the original matrix).

It should be noted that the triangular cofactors calculated in the last two operators belong to the same commutative domain as the elements of the original matrix, and the diagonal matrices comprise some of the elements from the quotient field of this domain. These matrix algorithms are described in [12–16].

## *5.5. Systems of Linear Differential Equations with Constant Coefficients*

This is another area where MathPartner outperforms both Mathematica and Maple. It solves large systems of linear differential equations with constant coefficients under given initial conditions. For example,

$$S = \mathbf{systLDE}(\mathbf{d}(y,t) + \mathbf{d}(x,t) - x = \mathbf{exp}(t),$$
$$2\mathbf{d}(y,t) + \mathbf{d}(x,t) + 2x = \mathbf{cos}(t))$$

defines the system of differential equations

$$S = \begin{cases} y'_t + x'_t - x = e^t, \\ 2y'_t + x'_t + 2x = cos(t). \end{cases}$$

$$J = \mathbf{initCond}(\mathbf{d}(x,t,0,0) = 0, \mathbf{d}(y,t,0,0) = 0)$$

is the specification of initial conditions.
Here,
- $\mathbf{d}(f,t)$ is the derivative *f* with respect to the variable *t*,
- $\mathbf{d}(f,t,k)$ is the *k*th derivative *f* with respect to *t*,
- $\mathbf{d}(f,t,k,t_0)$ is the *k*-th derivative *f* with respect to *t* in the point $t_0$ or $f(t_0)$ for $k = 0$.

The algorithms for solving systems of differential equations are described in [17–19].

*5.6. Noncommutative Objects*

All symbolic variables are regarded as commutative ones, except for those whose names are capitalized and begin with the symbol *backslash* (for example, $\backslash A$, $\backslash Omega$). Hence, for $a*b - b*a$, the result is zero, while, for \A*\B-\B*\A, the result is \A*\B-\B*\A,.

*5.7. Computations in Tropical Algebras*

An important feature of the MathPartner system is the support of symbolic computations in tropical algebras.

The following idempotent semifields are defined: ZMaxPlus, ZMinPlus (on the set Z), RMaxPlus, RMinPlus, RMaxMult, RMinMult (on the set R), R64MaxPlus, R64MinPlus, R64MaxMult, and R64MinMult (on the set R64).

The following idempotent semirings are defined: ZMaxMin, ZMinMax, ZMaxMult, ZMinMult (on the integer set Z), RMaxMin, RMinMax (on the set R), R64MaxMin, and R64MinMax (on the set R64).

There are, in total, 18 algebras of different types. Below is an example of a simple problem in ZMaxPlus:

```
SPACE = ZMaxPlus[];
a=2; b=9; c=a+b; d=a*b; \print(c,d);
```
The result is c=9; d=11.

A unary *closure* operator is defined as $\mathbf{closure}(a)$ (where $a$ is a number or a matrix) and evaluates the expression $1 + a + a^2 + a^3 + \ldots$.

In tropical semifields, there are following operators:
- $\mathbf{solveLAETropic}(A,b)$ solves a system of equations $Ax = b$;
- $\mathbf{BellmanEquation}(A,b)$ solves a system of Bellman equations $Ax + b = x$;
- $\mathbf{BellmanInequality}(A,b)$ solves a system of Bellman inequalities $Ax + b \leq x$.

*5.8. Graph Problems*

When, for a graph, a matrix $A$ of distances between adjacent vertices is defined,
- $\mathbf{searchLeastDistances}(A)$ finds the least distances between all vertices, and
- $\mathbf{findTheShortestPath}(A,i,j))$ finds the shortest path between two vertices $i$ and $j$.

# 6. PROCEDURES, FUNCTIONS, AND CONTROL OPERATORS

The body text of a program can be preceded by *procedures* and *functions*. Their declaration begins with the functional word \**procedure** followed by a name and (if any) arguments enclosed in parentheses. The function must have a return operator (\**return** objectName), while the procedure can have an exit operator (\**return**).

Procedures and functions have the following syntax:
- \**procedure***proc*1($arg$1, $arg$2,..){ $op_1$; $op_2$;...}

All functional words, operator names, and function names are preceded by the symbol "\". Hereinafter, for simplicity, we omit this symbol and write the operators in bold.

Like C and Java, the Mathpar supports the following *control* operators:
- $\mathbf{if}()\{\}\mathbf{else}\{\}$ is the branching operator;
- $\mathbf{while}()\{\}$ is the pre-test loop;
- $\mathbf{for}(;;)\{\}$ is the for-loop.

# 7. VISUALIZATION

The system has a full set of standard tools for constructing 2D and 3D plots and animations.

## 7.1. 2D Plots of Functions

The operators for constructing 2D plots of functions are as follow:
- **plot**$(f,[a,b])$ constructs a plot for an explicit function $f = f(x)$, $x \in (a,b)$;

**paramPlot**$([X,Y],[a,b])$ constructs a plot for a parametric function $X = X(x)$, $Y = Y(x)$, $x \in (a,b)$;

**tablePlot**() constructs a plot for a function given by a value table (all points are successively connected by straight lines);

- **pointsPlot**() constructs a plot for a function given by a value table (points are not connected) and allows a text label to be placed near any point;

**plotText**() creates text labels (one can specify the position of the label, font size, and the slope of the label);

- **showPlots**$([p1, p2,..])$ depicts several plots on one image.

The system allows the users to select certain segments of the plots, animate them, and change many of their parameters. See the user manual for more details.

## 7.2. Surface Images

Modern graphics cards make it possible to construct high-quality images on the user computer. For this purpose, they must be provided with the grid of the surface to be visualized, along with some additional parameters. However, the user may not have such a graphics card; in this case, the whole image is constructed on the server.

The Mathpar language supports two types of operators for constructing 3D plots.

1. Image is constructed on the side of the server:

- **plot3d**$(f,[x_0, x_1, y_0, y_1])$ constructs a plot for an explicitly defined surface $f = f(x,y)$, $x \in (x_0, x_1)$, $y \in (y_0, y_1)$;
- **paramPlot**$([X,Y,Z],[u_0, u_1, v_0, v_1])$ constructs a plot for a parametrically defined surface $X = X(u,v)$, $Y = Y(u,v)$, $Z = Z(u,v)$, $u \in (u_0, u_1)$, $v \in (v_0, v_1)$.

2. Image is constructed on the side of the client:

- **explicitPlot3d**$(Z)$ constructs a plot for an explicitly defined surface $Z = Z(x,y)$, $x \in (x_0, x_1)$, $y \in (y_0, y_1)$;
- **parametricPlot3d**$(X,Y,Z)$ constructs a plot for a parametrically defined surface $X = X(u,v)$, $Y = Y(u,v)$, $Z = Z(u,v)$, $u \in (u_0, u_1)$, $v \in (v_0, v_1)$;
- **implicitPlot3d**$(F)$ constructs a plot for an implicitly defined surface $F(x,y,z) = 0$, $x \in (x_0, x_1)$, $y \in (y_0, y_1)$, $z \in (z_0, z_1)$.

# 8. CLUSTER COMPUTATIONS

The MathPartner is capable of connecting to a computer cluster and enables parallel computations with certain operators on the cluster. The corresponding parallel computing algorithms are described in [20, 21].

For cluster computations, the following constant cluster parameters should be specified: TOTALNODES is the total number of cluster nodes allocated for computations, PROCPERNODE is the number of MPI processes to be run on one node, CLUSTERTIME is the maximum runtime (in minutes) of a program, and MAXCLUSTERMEMORY is the amount of memory allocated to the Java Virtual Machine. The following parallel operators are implemented.

For classical spaces (Z[x]):
- **adjointDetPar()** calculates the matrix inverse (adjoint matrix and determinant).

For tropical spaces (R64MaxPlus[x]):
- **BellmanEquationPar()** solves a system of Bellman equations $Ax + b = x$;
- **BellmanInequalityPar()** solves a system of Bellman inequalities $Ax \leq x$ and $Ax + b \leq x$.